# Reconfigurable directional couplers and junctions optically induced by nondiffracting Bessel beams

Zhiyong Xu, Yaroslav V. Kartashov, Victor A. Vysloukh,* Lluis Torner

*ICFO-Institut de Ciencies Fotoniques, and Department of Signal Theory and Communications, Universitat Politecnica de Catalunya, 08034 Barcelona, Spain*

We put forward the concept of reconfigurable structures optically induced by mutually incoherent nondiffracting Bessel beams in Kerr-type nonlinear media. We address collinear couplers and X-junctions, and show how the switching properties of such structures can be tuned by varying the intensity of the Bessel beams.

Since its theoretical prediction[1] and experimental observation,[2,3] guiding of probe beams by bright and dark spatial solitons has been studied comprehensively due to its potential applications to all-optical switching. The canonical example is the nonlinear directional coupler constituted by several parallel soliton-induced waveguides,[4,5] where an input signal periodically switches between the output channels.[6-9] Another important example is the X-junction, formed with several intersecting soliton beams.[10,11] The potential of such structures relies on the possibility of tuning the device characteristic by changing the properties of solitons employed to induce the switching structure. Here we suggest a different route to create reconfigurable devices for all-optical switching, based on the concept of optical lattices.

It was demonstrated recently[12-14] that reconfigurable honeycomb optical lattices can be induced with several interfering plane waves, propagating in linear regime, while vectorial interactions in slow Kerr-type media can be used to guide soliton beams in the lattice. Lattices with a radial symmetry induced by nondiffracting Bessel beams are also possible, and open a wealth of new opportunities.[15,16] For example, the nondiffracting beams can induce well-defined guiding channels that can trap and steer soliton beams.

Good approximation to Bessel beams can be generated experimentally.[17] Here we address directional couplers and X-junctions induced by several mutually incoherent parallel or intersecting Bessel beams and show the various soliton switching scenarios accessible with such optically induced and, hence, tunable devices. Notice that, in contrast to Refs [4,5], which address the switching of linear beams in soliton-induced waveguides, here we show that guiding structures created with linear beams lead to soliton control.

We consider the propagation of a light beam along the $z$ axis in a nonlinear cubic Kerr-type medium with an optically induced modulation of the refractive index. The evolution of the complex amplitude $q$ of the field can be described by the equation

$$i\frac{\partial q}{\partial \xi} = -\frac{1}{2}\left(\frac{\partial^2 q}{\partial \eta^2} + \frac{\partial^2 q}{\partial \zeta^2}\right) - q|q|^2 - pR(\eta,\zeta,\xi)q, \qquad (1)$$

where $\eta,\zeta$ and $\xi$ stand for the transverse and the longitudinal coordinates scaled to the beam width and diffraction length, respectively. We assume that the refractive-index modulation is induced by several incoherent zero-order Bessel beams, so that the refractive-index profile features the total intensity of the beams. The parameter $p$ is proportional to the modulation depth, and in the simplest case of two incoherent Bessel beams the refractive index modulation is described by $R(\eta,\zeta,\xi) = J_0^2[(2b_{lin})^{1/2}((\eta+\eta_0-\alpha\xi)^2+\zeta^2)^{1/2}] + J_0^2[(2b_{lin})^{1/2}((\eta-\eta_0+\alpha\xi)^2+\zeta^2)^{1/2}]$. The scaling parameter $b_{lin}$ defines the radii of rings of Bessel beams and we take it small enough to ensure that the width of the central beam core ($\sim 20~\mu$m) largely exceeds the wavelength, $2\eta_0$ is the initial separation between beam centers, and $\alpha$ defines the head-on intersection angle. Equation (1) admits several conserved quantities including the power or energy flow $U = \int_{-\infty}^{\infty}\int_{-\infty}^{\infty} |q|^2\, d\eta d\zeta$. We study the propagation of ground-state solitons launched parallel to one of the guiding channels of the structure. When searching for input soliton profiles we assume that only one guiding channel is present. The soliton profile has the form $q(\eta,\zeta,\xi) = w(\eta,\zeta)\exp(ib\xi)$, where $w(\eta,\zeta)$ is a real function and $b$ is a real propagation constant. As $b \to \infty$ the soliton width goes to zero while its power approaches the critical value $U_{cr} \approx 5.85$ given by power of unstable soliton in uniform cubic medium. Here we consider solitons with powers well below critical one, when coupling between Bessel waveguides is considerable. In the case of X-junctions, we impose the phase tilt $\exp(i\alpha\eta)$ onto the input solitons. Our primary aim in this Letter is to study the impact of the modulation depth $p$, the intersection angle $\alpha$, and the input soliton power $U$ on possible switching scenarios.

First we address properties of reconfigurable directional couplers induced with collinear incoherent Bessel beams ($\alpha = 0$). Because of the overlap of the soliton tails guided by neighboring Bessel channels, energy is exchanged periodically between the channels upon propagation. The rate of the energy exchange is given by the overlap integral that increases with decrease of separation between waveguides or with increase of refractive index modulation depth. Fig. 1 shows different soliton switching scenarios in two-core (a)-(c) and three-core (d)-(f) couplers for different input powers. At small powers one achieves almost total energy transfer from the input guiding channel into neighboring channels at coupling length $\xi = L_c$ (Figs 1(a) and 1(d)). Note that in the case of three-core coupler the energy is equally redistributed between two output channels. Similar phenomena occur with more complicated Bessel beam arrays arranged into ring configurations, when the energy flow of soliton launched into single channel is redistributed between all other channels at the output. Because of the periodic character of energy exchange the input field distribution is completely restored at $\xi = 2L_c$. At the critical power level, the energy is equally distributed between all channels of the coupler (Figs 1(b) and 1(e)). In this case the coupling length $L_c$ diverges. Finally at high powers there is no energy transfer into neighboring channels (Figs 1(c) and 1(f)). Thus, the optically induced coupler behaves as the sought-after nonlinear directional coupler.

The variation of the coupling length and the normalized transmission efficiency versus the input power for two-core coupler are shown in Fig. 2(a) and 2(b), respectively. The transmission efficiency is defined as the ratio of energy concentrated in the output channel to that concentrated in the input channel. The coupling length increases and transmission efficiency decreases as the input power approaches the critical value corresponding to equal energy distribution in all channels. To stress the reconfigurability afforded by optically induced couplers, in Figs 2(c) and 2(d) we plot the coupling length and the transmission efficiency versus the modulation depth $p$, at fixed input power. The modulation depth can be directly controlled by intensities of Bessel beams forming the coupler. Since at fixed power the soliton supported by single channel becomes narrower with increase of $p$, the transmission efficiency decreases above the critical modulation depth and all switching scenarios from total to negligible energy transfer can be achieved.

Second we address properties of X-junctions created with two intersecting Bessel beams ($\alpha \neq 0$). The initial separation between guiding channels is high enough so that at the initial stage of propagation soliton launched into right channel of the junction is almost unaffected by the presence of left channel. We calculate the transmission efficiency defined as the ratio of output power concentrated in the central core of left channel at $\xi = 2\eta_0/\alpha$ to the input power concentrated in the central core of the right channel at $\xi = 0$. Figure 3(a) shows that the intersection of two incoherent Bessel beams produces an area of locally increased refractive index that is elongated along $\xi$ axis. The smaller the intersection angle the longer the area of locally increased refractive index. When soliton from right channel enters this area it is bounced back, as it is visible in Figs 3(b)-3(d). Depending on the intersection angle (i.e., the length of intersection area) the soliton remains in the input channel (Fig. 3(b)), splits into two beams (Fig. 3(c)), or experiences total switching into the left channel (Fig. 3(d)). The potential applications of this effect for angle-controlled soliton switching are clearly visible. The normalized transmission efficiency versus intersection angle is shown in Fig. 4(a). Almost 100% switching contrast can be achieved in such X-junction. With further growth of $\alpha$ the transmission efficiency monotonically decreases. Small angles are not shown in Fig. 4(a), because in this case small modifications of the input angle result in drastic changes in switching dynamics and fast oscillations of the curve $T(\alpha)$ at $\alpha \to 0$. We found that switching with high contrast can be achieved by tuning the depth of refractive index modulation $p$ (Fig. 4(b)). There exists an optimal value of $p$ yielding almost total soliton switching into left channel at fixed input power and intersection angle. This is another confirmation of the potential of reconfigurable guiding structures induced with arrays of Bessel beams for all-optical soliton control.

In conclusion, we have shown that reconfigurable directional couplers and X-junctions induced with nondiffracting Bessel beams afford a variety of opportunities for soliton switching. The key feature we put forward is the possibility to control the device dynamics by adjusting the properties of the Bessel beams used to induce the coupler.

*Also with Departamento de Fisica y Matematicas, Universidad de las Americas-Puebla, Santa Catarina Martir, 72820, Puebla, Mexico.

# References with titles

# References without titles

# Figure captions

Figure 1. Switching scenarios in two- (a)-(c) and three-core (d)-(f) optically-induced couplers. Output intensity distributions are shown at $\xi \approx L_{\mathrm{c}}$. White contour lines are to help the eye and show positions of optically-induced channels. In the two-core coupler, the soliton is launched into left channel; in the three-core coupler it was launched into right channel. Input power $U = 1.56$ (a), 1.67 (b), 3.1 (c), 1.56 (d), 2.15 (e), and 2.68 (f). Parameters $p = 5$, $2\eta_0 = 3$, $b_{lin} = 10$.

Figure 2. (a) Coupling length and (b) normalized transmission efficiency versus input power at $p = 5$, $\eta_0 = 1.5$. (c) Coupling length and (d) normalized transmission efficiency versus modulation depth at $U = 2$, $\eta_0 = 1$. In both cases $b_{lin} = 10$.

Figure 3. (a) X-junction made by intersecting incoherent Bessel beams at $\alpha = 0.5$. Different soliton propagation scenarios at $\alpha = 0.5$ (b), $\alpha = 0.6$ (c), and $\alpha = 0.75$ (d). Intensity distribution are shown at $\zeta = 0$. In all cases $p = 5$, $U = 2$, $2\eta_0 = 6$, $b_{lin} = 2$.

Figure 4. (a) Normalized transmission efficiency versus intersection angle at $p = 5$, $U = 2$. (b) Normalized transmission efficiency versus modulation depth at $\alpha = 0.75$, $U = 2$. Parameters $2\eta_0 = 6$, $b_{lin} = 2$.

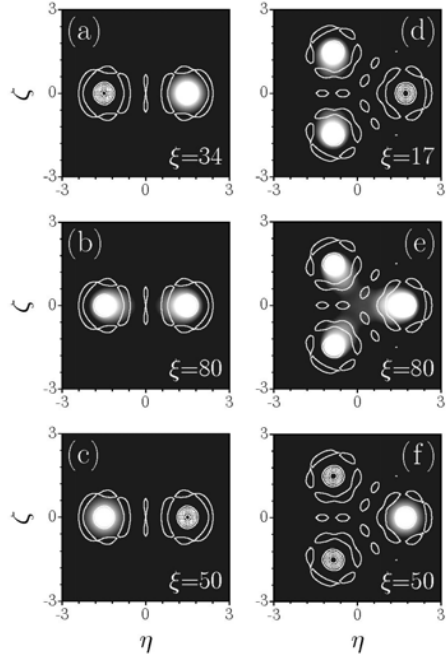

Figure 1. Switching scenarios in two- (a)-(c) and three-core (d)-(f) optically-induced couplers. Output intensity distributions are shown at $\xi \approx L_c$. White contour lines are to help the eye and show positions of optically-induced channels. In the two-core coupler, the soliton is launched into left channel; in the three-core coupler it was launched into right channel. Input power $U = 1.56$ (a), 1.67 (b), 3.1 (c), 1.56 (d), 2.15 (e), and 2.68 (f). Parameters $p = 5$, $2\eta_0 = 3$, $b_{lin} = 10$.

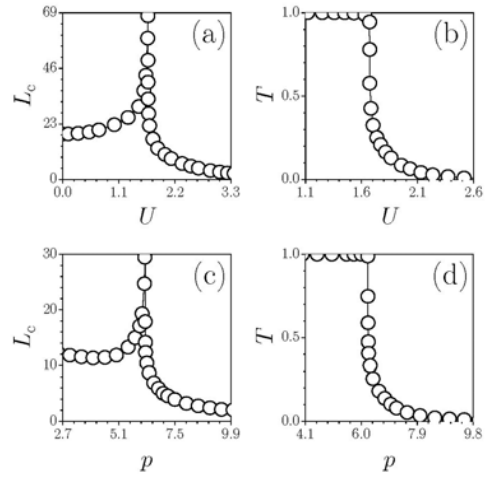

Figure 2.  (a) Coupling length and (b) normalized transmission efficiency versus input power at $p = 5$, $\eta_0 = 1.5$. (c) Coupling length and (d) normalized transmission efficiency versus modulation depth at $U = 2$, $\eta_0 = 1$. In both cases $b_{lin} = 10$.

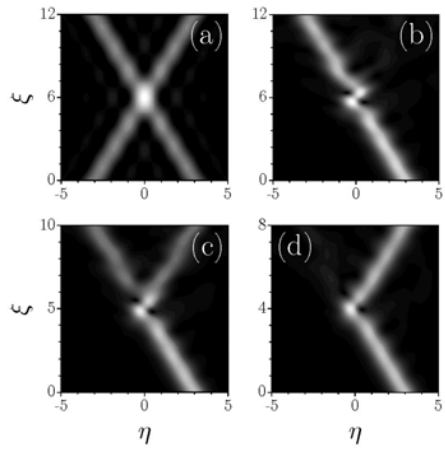

Figure 3. (a) X-junction made by intersecting incoherent Bessel beams at $\alpha = 0.5$. Different soliton propagation scenarios at $\alpha = 0.5$ (b), $\alpha = 0.6$ (c), and $\alpha = 0.75$ (d). Intensity distribution are shown at $\zeta = 0$. In all cases $p = 5$, $U = 2$, $2\eta_0 = 6$, $b_{lin} = 2$.

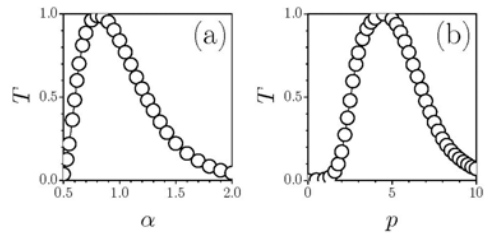

Figure 4. (a) Normalized transmission efficiency versus intersection angle at $p = 5$, $U = 2$. (b) Normalized transmission efficiency versus modulation depth at $\alpha = 0.75$, $U = 2$. Parameters $2\eta_0 = 6$, $b_{lin} = 2$.